\documentclass[10pt,twocolumn,letterpaper]{article}

\usepackage{cvpr}
\usepackage{times}
\usepackage{epsfig}
\usepackage{graphicx}
\graphicspath{ {./images/} }
\usepackage{amsmath}
\usepackage{amssymb}
\usepackage{subfigure}
\usepackage{amsmath}

\usepackage[breaklinks=true,bookmarks=false]{hyperref}

\cvprfinalcopy 

\def\cvprPaperID{****} 

\begin{document}

\title{Emotion Recognition in Audio and Video Using Deep Neural Networks}

\author{Mandeep Singh\\
SCPD, Stanford University\\
Stanford, CA\\
{\tt\small \href{https://www.linkedin.com/in/smandeep/}{msingh13@stanford.edu}}
\and
Yuan Fang\\
ICME, Stanford University\\
Stanford, CA\\
{\tt\small \href{mailto:yuanfy@stanford.edu}{yuanfy@stanford.edu}}
}

\maketitle

\begin{abstract}

Humans are able to comprehend information from multiple domains for e.g. speech, text and visual. With advancement of deep learning technology there has been significant improvement of speech recognition. Recognizing emotion from speech is important aspect and with deep learning technology emotion recognition has improved in accuracy and latency. There are still many challenges to improve accuracy. In this work, we attempt to explore different neural networks to improve accuracy of emotion recognition. With different architectures explored, we find (CNN+RNN) + 3DCNN multi-model architecture which processes audio spectrograms and corresponding video frames giving emotion prediction accuracy of 54.0\% among 4 emotions and  71.75\% among 3 emotions using IEMOCAP\cite{IEMOCAP} dataset.

\end{abstract}

\section{Introduction}

Emotion recognition is an important ability for good interpersonal relations and plays an important role in an effective interpersonal communications. Recognizing emotions, however, could be hard; even for human beings, the ability of emotion recognition varies among persons.

The aim of this work is to recognize emotions in audios and audios+videos using deep neural networks. In this work, we attempt to understand bottlenecks in existing architecture and input data, and explored novel ways on top of existing architectures to increase emotion recognition accuracy.

The dataset we used is IEMOCAP\cite{IEMOCAP}, which contains 12 hours audiovisual data of 10 people(5 females, 5 males) speaking in anger, happiness, excitement, sadness, frustration, fear, surprise, other and neutral state. 

Our work mainly consists of two stages. First, we build neural networks to recognize emotions in audios by replicating and expanding upon the work of \cite{inproceedings}. The input of the models are the audio spectrograms converted from the audio of an actor speaking a sentence, and the models give one output which is the emotion the actor has when saying that sentence. The  models  only  predict  one  of  the  four  different emotions, e.g.  happiness, anger, sadness, and neutral state, which were chosen for comparison with \cite{inproceedings}. The deep learning architectures we explored were CNN, CNN + RNN, CNN + LSTM. 

After achieving a comparably good accuracy on audios comparing with \cite{inproceedings}, we build models which predict emotions using audio spectrogram and video frames in a video, since we believe video frames contain additional emotion-related information that can help us to achieve a better emotion prediction performance. The inputs of the models are the audio spectrogram and video frames, which are converted and extracted from the sound and images of a video recording an actor speaking one sentence. The output of the models is still one of the four selected emotions mentioned above. Inspired by the work of \cite{DBLP:journals/corr/TorfiIND17}, we explore the model made of two sub-networks; the first sub-network is a 3D CNN which takes in the video frames, and the second one is a CNN+RNN which takes in the audio spectrogram, and the last layer of the two sub-networks are concatenated and followed by a fully connected layer that output the prediction.

The metric we use for evaluation is the overall accuracy for both the audio and audio+video models.




\begin{figure}[t]
\begin{center}
\fbox{
   
   \includegraphics[width=8cm,height=6cm]{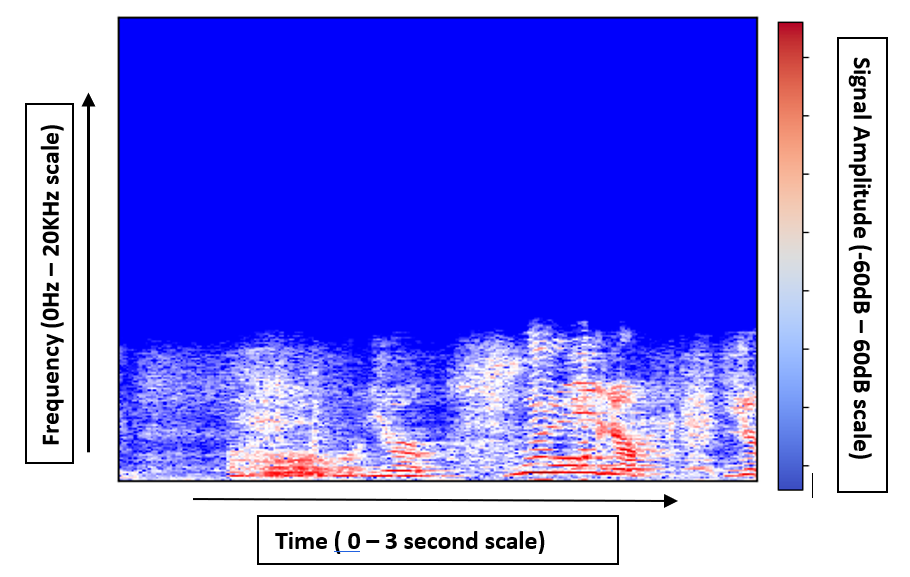}
   \centering 
   }
\end{center}
   \caption{Example of audio spectrogram of anger emotion. Original time scale without noise cleanup.}
\label{fig:long}
\label{fig:onecol1}
\end{figure}

\section{Related Work}


Emotion recognition is an important research area that many researchers work on in recent years using various methods. Using speech signals\cite{kwon2003emotion}, facial expression\cite{gouta2000emotion}, and physiological changes\cite{kim2008emotion} are some of the common approaches researchers arise to approach the emotion recognition problem. In this work, we will use audio spectrograms and video frames to do emotion recognition.

It has been shown, emotion recognition accuracy can be improved with statistical learning of low-level features (frequency \& signal power intensity) by different layers of deep learning network. Mel-scale spectrograms for speech recognition was demonstrated to be useful in \cite{deng_2014}. There has been state of the art speech recognition method that uses linearly spaced audio spectrograms as described in \cite{AmodeiABCCCCCCD15} \cite{HannunCCCDEPSSCN14}. Our work related to emotion recognition using audio spectrogram follows the approach described in \cite{inproceedings}. Audio spectrogram is an image of audio signal which consists of 3 main components namely: 1. Time on x-axis. 2. frequency on y-axis. 3. power intensity on the colorbar scale which can be in decibels(dB) as shown in Fig. 1. \cite{sahu} covers machine learning methods to extract temporal features from the audio signals. The goodness in machine learning models is, it's training \& prediction latency is good but, prediction accuracy is low. The CNN model that uses audio spectrograms to detect emotion has better prediction accuracy compared to machine learning model.

Comparing the CNN network used in \cite{inproceedings} \& \cite{DBLP:journals/corr/TorfiIND17} for training using audio spectrograms, \cite{inproceedings} uses wider kernel window size with zero padding while \cite{DBLP:journals/corr/TorfiIND17} uses smaller window size and no zero padding. With wider kernel window size we are able to see larger vision of the input which allows for more expressive power. In order to avoid loosing features use of zero padding becomes important. The zero padding decreases as the number of CNN layers increase in the architeture used in \cite{inproceedings}. \cite{DBLP:journals/corr/TorfiIND17} avoids adding zero padding in order to not consume extra virtual zero-energy coefficients which are not useful in extracting local features. One drawback that we see in \cite{DBLP:journals/corr/TorfiIND17} is that it does not compare performance between audio model \& audio+video model being used. One goodness observed in \cite{DBLP:journals/corr/TorfiIND17} is that it does not do noise removal from the audio input data while \cite{inproceedings} uses noise removal techniques in the audio spectrogram before training the model.


To achieve better prediction accuracy, a natural progression of emotion recognition using audio sprectrogram is to include facial features extracted from the video frames. \cite{DBLP:journals/corr/abs-1902-01019} \& \cite{article_facial_video} implements facial emotion recognition using images and video frames respectively but, without audio. \cite{DBLP:journals/corr/TorfiIND17} \& \cite{DBLP:journals/corr/abs-1807-00230} implements neural network architecture which processes audio spectrogram \& video frames to recognize emotion. Both \cite{DBLP:journals/corr/TorfiIND17} and  \cite{DBLP:journals/corr/abs-1807-00230} implement a self-supervised model for cooperative learning of audio \& video models on different dataset.  \cite{DBLP:journals/corr/abs-1807-00230} further does a supervised learning on the pre-trained model to do classification. The model come up by \cite{DBLP:journals/corr/TorfiIND17} and  \cite{DBLP:journals/corr/abs-1807-00230} are very similar; both are two-stream models that contains one part for audio data, and one part for video data. The only difference is the way the kernel size, layer number, input data dimension are set. These hyperparamters are set differently because their input data is different.\cite{DBLP:journals/corr/TorfiIND17} tends to use smaller input size, and kernel size because its input images only capture mouth, which doesn't contain as much information as the image which captures the movement of a person used in  \cite{DBLP:journals/corr/abs-1807-00230}.


\section{Dataset \& Features}

\subsection{Dataset}
The dataset we use is IEMOCAP \cite{IEMOCAP} corpora as it is the best known comprehensibly labeled public corpus of emotion speech by actors.\cite{lee2015} uses this IEMOCAP dataset to generate state of the art results at the time. IEMOCAP contains 12 hours audio and visual data of conversations of two persons (1 female and 1 male for one conversation, and there are 5 females and 5 males in total), where each sentence in conversations is labelled with one emotion--anger, happiness, excitement, sadness, frustration, fear, surprise, other or neutral state.
\subsection{Data pre-processing}
\subsubsection{Audio Data Pre-processing}

IEMOCAP data corpus contain audio wav files with various time length and with marking of actual emotion label  for corresponding time segment. The audio wav files in IEMOCAP are generated at a sample rate of 22KHz. The audio spectrogram is extracted from the wav file using librosa\footnote {https://librosa.github.io/librosa/index.html} python package with a sample rate of 44KHz. 44KHz sample rate was used because as per Nyquist-Shannon sampling theorem\footnote{https://en.wikipedia.org/wiki/Nyquist\%E2\%80\%93Shannon\_sampling\_theorem} in order to fully recover a signal the sampling frequency should at least be twice the signal frequency. The audio signal frequency ranges from 20Hz to 20KHz. Hence, 44KHz sampling rate is commonly used rate for sampling. The spectrograms were generated in 2 segments which are: 1. Original time length of utterance of sentence or emotion. 2. Clip each utterance of sentence into 3 second clips. Another data segmentation that was done is with noise cleanup and without noise cleanup. We have named these segmentation as DSI, DSII, DSIII \& DSIV. This data segmentation is summarized in Table 1. Model training is done on these data segments separately.
\begin{table}
\begin{small}
\begin{center}
\begin{tabular}{|l|c|c|}
\hline
Dataset segmentation Type & Noise Cleanup & Name\\ 
\hline\hline
Original time length of utterance  & No & DS I\\
Clip each utterance into 3 second clips & No & DS II\\
Original time length of utterance  & Yes & DS III\\
Clip each utterance into 3 second clips & Yes & DS IV\\
\hline
\end{tabular}
\end{center}
\caption{Segmentation of input data generation.}
\end{small}
\end{table}

\begin{figure}[t]
\begin{center}
\fbox{
   
   \includegraphics[width=8cm,height=6cm]{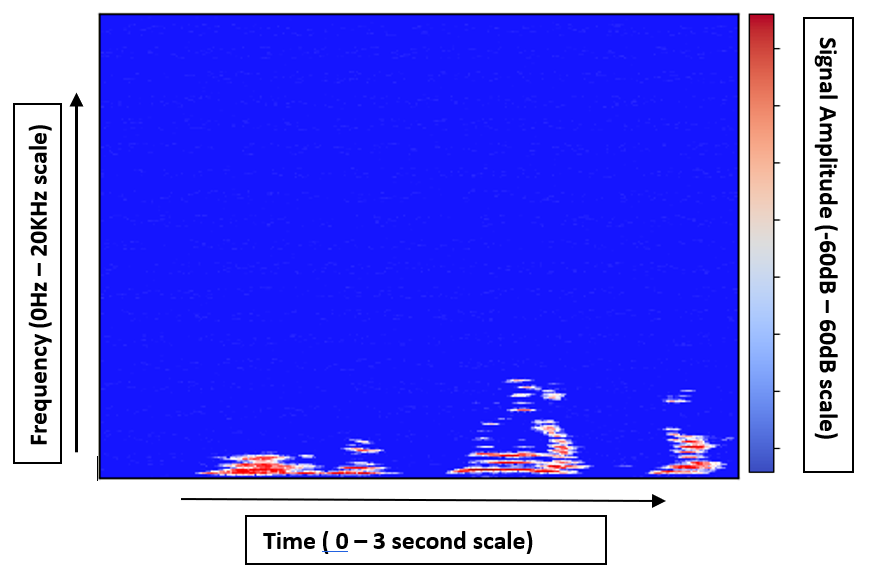}
   \centering 
   }
\end{center}
   \caption{Example of audio spectrogram of anger emotion. 3 sec audio clip with noise cleanup. Compare with Fig.1}
\label{fig:long1}
\label{fig:onecol}
\end{figure}

In order to get rid of the background noise we applied bandpass filter\footnote{\small{https://timsainburg.com/noise-reduction-python.html}} \footnote{\small{https://github.com/julieeF/CS231N-Project/blob/master/load\_single\_wav\_file.py}} between 1Hz to 30KHz. Denoising or noise cleanup of the input audio signal for data augmentation is also followed by \cite{AmodeiABCCCCCCD15}. Sentence utterances that are less than 3 second are padded with noise  to maintain uniformity in noise frequency and noise amplitude w.r.t noise in other parts of the signal. Initially zero padding was also experimented with to have 3 second time scale and then noise is added with signal to noise ratio (SNR) of 1 through out the signal time length but, this resulted in distorting the original audio signal. The resulting signal is then denoised. Denoising helps in making the frequency, time scale and amplitude features of the input audio signal to be more visible in the hopes of getting better prediction accuracy per emotion. All the audio spectrograms are generated with same colorbar intensity scale (+/- 60dB) to maintain uniformity of the spectrogram across the board among different emotions. This is similar to normalization of data.  As seen in Fig. 2 after denoising only the signal that contains actual information remain with high power intensity or signal amplitude. Other regions in the spectrogram remains with low power intensity relative to where there is actual signal of interest. Compared to Fig. 1 where some signal intensity is observed throughout the time scale, which is actually the noise. The spectrogram images generated are of size 200x300 pixels.

Total count of 3 second audio spectrograms among 4 different emotions is summarized in Table. 2. As observed the happy emotion count is significantly low. So we duplicated the happy data to reach total count of 1600. Similarly anger emotion count was also duplicated. Sad \& Neutral data count was reduced to match to 1600 data points for each emotion. A total images of 6400 is used for training the model. Data balance is crucial for the model to train well. 400 images from each emotion is used for model validation purpose. The images used for validation are never part of training set.

At first, we started off with audio spectrograms that contains xy axis and colorbar scale but, we removed the scales after learning that including axis \& scale could be contributing negatively to prediction accuracy.

To observe class accuracy improvement, input audio spectrograms were data augmented by cropping and rotation. Each image was cropped by 10 pixels from the top and resized back to 200x300 pixels.  This cropping is done to simulate frequency change in the emotion by small amount. Similarly, each image was also rotated by +/- 10 degrees. This rotation also simulates frequency change but it also shifts the time scale. Augmenting data that changes time scale is not preferred hence the rotation was done to a very small degree of 10 degrees. With cropping and rotation, total count of data used for training becomes 19200. The model training was done separately with original images and images with data augmentation for comparison.Horizontal flip of images were avoided as this means flipping the timescale and enacts a person speaking in reverse, which will lead to lower model prediction accuracy.

Model training on audio spectrogram that contains full time length, and not 3 second, was done separately. For given 3 second audio spectrogram, it was replaced with the full time length spectrogram, thus maintaining data count for balancing.

Visual analysis of around 100 audio spectrograms were done. It was observed that maximum frequency observed among all these spectrograms is around 8KHz. This means around 60\% of the spectrogram image is blue which does not carry any information from emotion perspective. All the input audio spectrograms were cropped from the top by 60\% and resized back to 200X300 pixels. An ideal method would be to generate spectrograms specifying fixed frequency scale if the frequency range is known prior.

\begin{table}
\begin{center}
\begin{tabular}{|l|c|}
\hline
Emotion & Count of data points\\ 
\hline\hline
Happy & 786\\
Sad & 1752 \\
Anger & 1458\\
Neutral & 2118 \\
\hline
\end{tabular}
\end{center}
\caption{Data count of each emotion.}
\end{table}

\subsubsection{Video Data Pre-processing}
Since our work also include implementing video model to see room for improvement in prediction accuracy of emotion recognition, we also did video data pre-processing. For the video data, we first clipped each video file into sentences according to how we processed audio files. This ensured that we are querying that part of the video file that corresponds to given audio spectrogram. Then we extracted 20 images per 3 second from each video avi file that corresponds to 3 sec audio spectrogram. The video contains both the actors in the frames hence the frames were cropped accordingly from left or right to only capture the actor whose emotion is being recorded. We then cropped the video frames further to cover the actor's face/head. The final resolution of video frames is 60x100. One limitation with the dataset is that, in the video the actors are not speaking facing the camera, therefore full facial expression corresponding to a given emotion are not visible.

While processing the extraction of audio spectrograms and video frames, it was observed that the memory usage on the machine was more than 12GB. This lead to machine crashes. Therefore, to extract data, each audio and video file was processed individually in batch. Python script\footnote{\small{https://github.com/julieeF/CS231N-Project/blob/master/load\_single\_Video.py}} was launched individually through unix shell script. 

\section{Methods \& Model Architecture}
In this section, we will talk about the models we have built for emotion recognition in audios in the 'Audio Models' subsection, and models for emotion recognition in audios+videos in 'Audio+Video Models' subsection.

\subsection{Audio Models}

\begin{figure}[t]
\begin{center}
\fbox{
   
   \includegraphics[width=8cm,height=10cm]{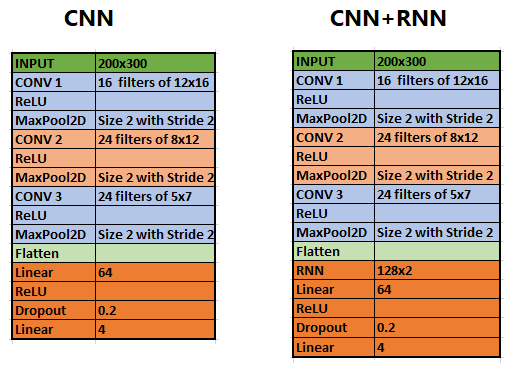}

   }
\end{center}
   \caption{Audio model architectures.}
   \label{fig:audiom}
\end{figure}

By replicating and expanding upon the network architecture used in \cite{inproceedings} we formulate three different models. The first model is a CNN model, which consists of three 2D convolutional layers and maxpooling layers followed by two fully connected layers, as shown in Fig.\ref{fig:audiom}. The second architecture we build is that we add a LSTM layer after the convolutional layers in the CNN model we have built, and we will call this model CNN+LSTM in this work. In the third model, we replace the LSTM layer with a vanilla RNN layer, and this model is named CNN+RNN in this work. A graph for the architecture of CNN+RNN is shown in Fig.\ref{fig:audiom}. The loss we use for training the model is the cross entropy loss. 
\begin{align}
    L_\textbf{cross entropy}=\frac{1}{N}\sum_{n=1}^N-\text{log}(\frac{\text{exp}(x_c^n)}{\sum_j \text{exp}(x_j)})
\end{align}
where N is the number of data in the dataset, $x_c^n$ is the true class's score of the n-th data point,  $x_j$ is the j-th class's score of the n-th input data. Minimizing the cross entropy loss will force our model to learn the emotion-related features from the audio spectrogram because when the loss will be minimum only when for a datapoint, the score of the true class is remarkably larger than the score of all other classes.


\begin{figure}[t]
\begin{center}
\fbox{
   
   \includegraphics[width=8cm,height=12cm]{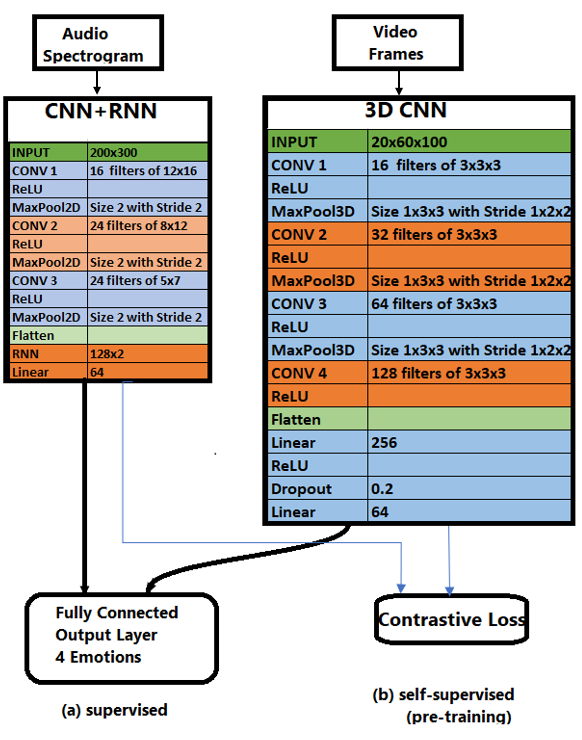}

   }
\end{center}
   \caption{Audio+Video model architectures.}
   \label{fig:videom}
\end{figure}


\subsection{Audio+Video Models}
Inspired by the work of \cite{DBLP:journals/corr/TorfiIND17}, our audio+video model is a two-stream network that consists of two sub-networks as shown in Fig.\ref{fig:videom}(a). The first sub-network is the audio model, which we choose to use the best-performing audio model we have built--CNN \& RNN as shown in Fig.\ref{fig:audiom}. The architecture of the first sub-network is the same as the audio model except that it dumps the original output layer in order to get high-level features of audio spectrograms, as shown in Fig.\ref{fig:videom}(CNN+RNN). The second sub-network is the video model, and it is made of four 3D convolutional layers, and three 3D maxpooling layers, followed by two fully connected layers, as shown in Fig.\ref{fig:videom}(3D CNN). Finally, the last layer of the two sub-networks are concatenated together, followed by one output layer, as shown in Fig.\ref{fig:videom}(a). 

We train this audio+video model using two different methods--semi-supervised training and supervised training. For semi-supervised training method, we first pre-train our model using video frames and audio spectrogram from the same video and from different videos, as shown in Fig.\ref{fig:videom}(b). This forces the model to learn the correlation between the visual and auditive elements of a video. The input of the pre-training process has three distinct types--positive (the audio spectrogram and video frames are from the same video); hard negative (the audio spectrogram and video frames are from different videos with different emotions); super hard negative (the audio spectrogram and video frames are from different videos with the same emotion). The loss function we use for pre-training is the contrastive loss. 
$$L_{\text{contrastive loss}}=\frac{1}{N}\sum_{n=1}^N L_1^n+L_2^n$$
where $$L_1^n=(y^n)\left\|f_v(v^n)-f_a(a^n)\right\|_2^2$$ 
$$L_2^n=(1-y^n)\text{max}(\eta-\left\|f_v(v^n)-f_a(a^n)\right\|_2,0)^2$$
N is the number of datapoints in the dataset, $v^n, a^n$ are the video frames and audio spectrogram of the n-the datapoint, $f_v, f_a $ are the video and audio sub-networks, $y_n$ is one if the video frames and audio spectrogram are from the same video, and zero otherwise.$\eta$ is the margin hyperparameter. $\left\|f_v(v^n)-f_a(a^n)\right\|_2$ should be small when the video frames and audio spectrogram are from the same video, and large when they come from different videos. Therefore, by minimizing the contrastive loss, the audio and video models are forced to output similar values when their inputs are from the same video, and very disctint values when they are not. This allows the model to learn the connection between audio and visual elements from the same video.

After pre-training is done, we do supervised learning on the pre-trained model where the input is the audio spectrogram and video frames of a video and output is the emotion predicted, as shown in Fig.\ref{fig:videom}(a). The loss of our model is the cross entropy, and the formula is the same as in Equation.1.

The second training method is that we do supervised training directly on the model without pre-training process.

   


\section{Experiments \& Results}

For model evaluation, prediction accuracy is the key metric used. For results comparison, the accuracy was compared with accuracy reported in \cite{inproceedings}. Since we balanced the data count therefore the overall accuracy and class accuracy as reported in \cite{inproceedings} are mathematically equal terms in our work. Our work aimed to achieve prediction accuracy of around 60\% considering 4 emotions.

We trained the model on all 4 segmentation of dataset generated and observed that results on data with original time scale and without noise cleanup gives the best accuracy. The results reported are based on this dataset. Spectrograms with noise removed theoretically sounds promising but it did not work due to 2 possible reasons. First, the algorithm used to remove noise reduces signal amplitude which may lead to some feature suppression. An algorithm that amplifies the signal back need to be explored. Some techniques for e.g. subtracting noise from the signal and multiplying final signal with a constant was explored but they all resulted in signal distortion. Secondly, having noise in the spectrogram simulates real scenario and during model training noise could indirectly acts as a regularizer. \cite{DBLP:journals/corr/TorfiIND17} also does not remove noise from the input audio spectrograms.

\subsection{Hyperparameters}
We started off with prediction on 4 emotions and most of the work, results and analysis is based off of these 4 emotions. Our validation accuracy did not go beyond 54.00\% and we saw overfitting during model training beyond this point. This lead us to experiment with various hyperparameters in the optimizer and in the network model layers for e.g. kernel size, size of input and output in each layer, dropout, batchnorm, data augmentation, l1 \& l2 regularization. 

Adam optimizer was used with learning rate of 1e-4 to train the model as this gave best accuracy. We experimented with 1e-3 \& 1e-5 and observed the model did not train well with these settings.
It was observed weight decay(parameter that controls l2 regularization) of 0.01 in Adam optimizer improved the accuracy by ~1\%. Weight decay values of 0.005 and 0.02 were also experimented with but, did not help.  All other parameters was kept default in the optimizer. 

Enabling l1 regularization, data augmentation of rotation and cropping, batchnorm resulted in no change or improvement in accuracy. This is possibly due to that the model learned all the features it can from the available data based on the model architecture.

Tuning of dropout probability was also experimented with and optimal value of 0.2 for the last fully connected layer and 0.1 for the dropout in RNN layer was obtained.

The input \& output dimensions in the audio network layers was doubled \& quadrupled which resulted in accuracy improvement of 1-2\%. Increasing the input output dimensions in the layers also resulted in high memory usage during model training. We attempted to extend this learning on the video network but due to limited memory of 12GB on the machine we were unable to carry out this experiment. This leads us to strongly believe that there is room for improvement that needs more experimentation on a machine with large memory. The accuracy improvement is also evident from different model architectures we used starting from CNN, CNN+RNN and CNN+RNN+3DCNN which actually is having more parameters in the model to learn features better.

80\% of total data points were used for training and rest for validation. Batch of 64 data points per iteration is used to train the model. Higher batch count resulted in long iteration time and high memory usage thus, 64 was picked.

Using normalization in image transformation with mean of [0.485, 0.456, 0.406] and standard deviation of [0.229, 0.224, 0.225] on all images improved accuracy by 0.37\%.

\subsection{Validation Set Accuracy}

Table \ref{fig:tab} summarizes the validation set accuracy obtained among different architectures. 

\begin{table}
\begin{small}
\begin{center}
\begin{tabular}{|l|c|c|c|}
\hline
Architecture & Accuracy(\%) & Data Aug. & Emotion \\
\hline\hline
CNN & 52.23 & No & H,S,A,N\\
CNN & 51.90 & Yes& H,S,A,N\\
CNN+LSTM & 39.77 & No& H,S,A,N\\
CNN+LSTM & 39.65 & Yes& H,S,A,N\\
CNN+RNN & 54.00 & No& H,S,A,N\\
CNN+RNN & 70.25 & No& S,A,N\\
CNN+RNN+3DCNN & 51.94 & No& H,S,A,N\\
CNN+RNN+3DCNN & 71.75 & No& S,A,N\\
\hline
\end{tabular}
\end{center}
\caption{Validation set accuracy over CNN, CNN+LSTM, CNN+RNN \& CNN+RNN+3DCNN among 4 and 3 different emotions. \small{H=Happy, S=Sad, A=Angry, N=Neutral}}

\label{fig:tab}
\end{small}
\end{table}

\subsection{Loss \& Classification Accuracy History on CNN+RNN+3DCNN}

\begin{figure}[t]
\begin{center}
\fbox{
   
   \includegraphics[width=8cm,height=6cm]{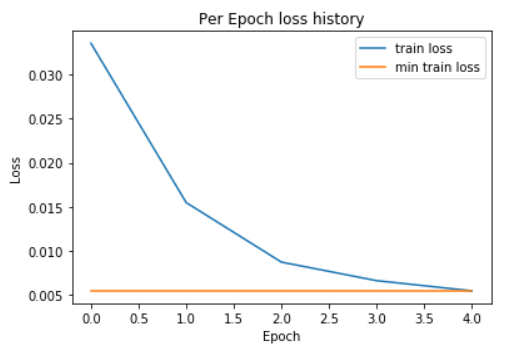}
   \centering 
   }
\end{center}
   \caption{Contrastive loss history curve on CNN+RNN+3DCNN.}
   \label{fig:long_loss4}
\end{figure}

Fig. \ref{fig:long_loss4} is the contrastive loss curve obtained during self supervision model training. We ran self supervision model with 5 epochs and 10 epochs separately and fed the learned weights in these 2 experiments into CNN+RNN+3DCNN model for classification training. We observed the self supervised model run with 5 epochs gave better classification accuracy by 0.5\% compared to the self supervised model run with 10 epochs. This could be attributed to overfitting of weights learned in self supervised model when run with 10 epochs.

\begin{figure}[t]
\begin{center}
\fbox{
   
   \includegraphics[width=8cm,height=6cm]{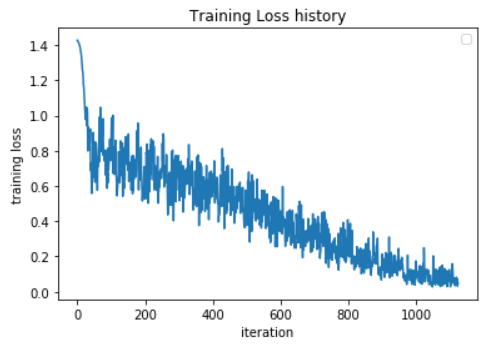}
   \centering 
   }
\end{center}
   \caption{Loss history curve on CNN+RNN+3DCNN. The curve is noisy because it is generated per iteration.}
   \label{fig:long_loss}
\end{figure}

Fig. \ref{fig:long_loss} is the softmax/cross entropy loss curve obtained on the best model which is CNN+RNN+3DCNN. Since the loss is reported per iteration hence it appears noisy but we observed that per epoch it is decreasing on logarithmic scale.

\begin{figure}[t]
\begin{center}
\fbox{
   
   \includegraphics[width=8cm,height=6cm]{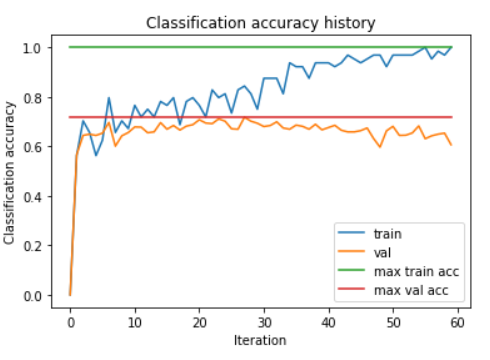}
   \centering 
   }
\end{center}
   \caption{Classification accuracy history on CNN+RNN+3DCNN after every 20 iterations for a total of 1200 iteration.}
   \label{fig:long_acc}
\end{figure}

Fig. \ref{fig:long_acc} is the classification accuracy history on best model which is CNN+RNN+3DCNN. We obtained best validation accuracy of 71.75\% considering 3 emotions(sad, anger, neutral).

\subsection{Confusion Matrix on CNN+RNN \& CNN+RNN+3DCNN}
\begin{figure}[t]
\begin{center}
\fbox{
   
   \includegraphics[width=8cm,height=6cm]{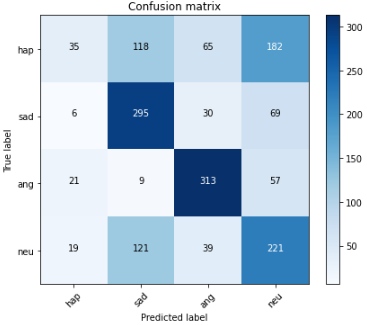}
   \centering 
   }
\end{center}
   \caption{Confusion matrix of true class vs. prediction in CNN+RNN.}
   \label{fig:long_conf}
\end{figure}

\begin{figure}[t]
\begin{center}
\fbox{
   
   \includegraphics[width=8cm,height=6cm]{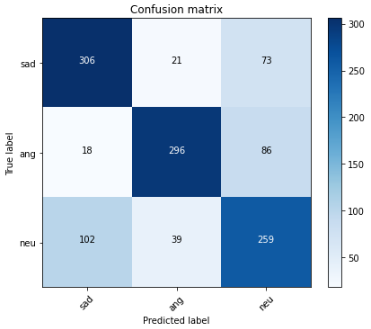}
   \centering 
   }
\end{center}
   \caption{Confusion matrix of true class vs. prediction in CNN+RNN+3DCNN.}
   \label{fig:long_conf_3_emo}
\end{figure}

Fig. \ref{fig:long_conf} is the confusion matrix obtained with CNN+RNN. From this confusion matrix we see only happy emotion is predicted poorly compared to other emotions. This led us to explore CNN+RNN \& CNN+RNN+3DCNN architecture only on 3 emotions (instead of 4) to understand if we do see performance improvement when switching from audio only inputs to audio+video inputs. Fig. \ref{fig:long_conf_3_emo} is the confusion matrix obtained with the best model which is CNN+RNN+3DCNN. 

\subsection{Results Analysis}

From Table \ref{fig:tab}, considering 4 emotions, we can see that CNN+RNN is the best performing architecture and data augmentation doesn't improve the accuracy. CNN does not work well comparing with CNN+RNN because CNN has the same architecture as the first few layers of CNN+RNN, and is comparably simple. Therefore, CNN+RNN will learn features of higher-level and performs better compared with CNN. For CNN+LSTM, it does have a more complex architecture, however, when we were tuning the hyperparams, we found out that accuracy improved slightly by increasing dropout probability in CNN+LSTM , indicating that CNN+LSTM could be overly complex for our dataset and training purpose. Also, adding model complexity requires more careful hyperparameters tuning, and since CNN+RNN is giving a relatively good performance compared with \cite{inproceedings}, we decided not to bother with adjusting CNN+LSTM. 

From Table \ref{fig:tab}, it also evident that CNN+RNN+3DCNN architecture which uses video frames along with audio spectrogram is the best considering 3 emotions but, the accuracy did not improve significantly to CNN+RNN. This is due to the fact that the cropping window to focus on the face/head to recognize facial emotion was large as the actors are not facing the camera and they moved during their speech. Auto detecting face/head with detection model and then cropping based on the bounding box would be ideal and accuracy is expected to increase significantly. Considering 4 emotions, CNN+RNN+3DCNN performed worse compared to CNN+RNN is because the model prediction accuracy for happy emotion itself is bad due its low data count, therefore when the video frames are used which are only using facial expression from the side only confuses the model to learn poorly.

Data augmentation does not increase the validation accuracy and even slightly makes the model perform worse could be due to that the image generated by cropping and rotation loses some emotion-related features, since it alters the frequency and time scale. Which is similar to altering the pitch of the audio and reversing the audio of a sentence, and could confuse the model.

From confusion matrix, we observed that happiness prediction is low compared to other emotions. One possible reason for this is, happiness data set count is very low compared to other emotions, and over-sampling by repetition the happiness data set is not enough. More dataset of happiness is expected to improve happiness prediction accuracy. 

Comparing our results with \cite{inproceedings}, we lag their class accuracy by 5.4\% but, comparing the overall accuracy considering 3 emotions our work achieved accuracy of 71.75\% which is better by 2.95\%.

\section{Conclusion/Future Work}

Our work demonstrated emotion recognition using audio spectrograms through various deep neural networks like CNN, CNN+RNN \& CNN+LSTM on IEMOCAP\cite{IEMOCAP} dataset. We then explored combining audio with video to achieve better performance accuracy through CNN+RNN+3DCNN. We demostrated that CNN+RNN+3DCNN performs better as it learns emotion features from the audio signal(CNN+RNN) and also learns emotion features from facial expression in video frames(3DCNN) thus complementing each other.

To further improve the accuracy of our model we plan to explore more and touch on various aspects. We want to explore more noise removal algorithms and generate audio spectrograms without noise in them. This work will help in analyzing if removing noise actually helps or it acts as a regularizer and we don't need to remove noise from the spectrograms. We also want to explore, if there are multiple people speaking how the model predicts the emotion in such scenarios.
Next, we want to explore auto cropping around the face/head from video frames. We strongly believe it will significantly improve the prediction accuracy.
As far as data augmentation is concerned, even though none of direct data augmentation methods proved to be useful but, adding signal with very low amplitude and varying frequency onto the speech signal and then generating audio spectrogram from the resulting signal will create unique data points and help in getting rid of model overfitting. 
If there were machines/GPUs with more memory we wanted to experiment with increasing input and output dimensions in each layer in the network to obtain optimal point. There is definitely a room to get better accuracy using this method. We then want to experiment with prediction latency among different models and there architecture size.
We also wanted to experiment more with CNN+LSTM network and fine tune it to see what is the best accuracy we can achieve with this model.
We did try transfer learning using ResNet18 but didn't achieve good results. Need to do more experimentation on how to transfer learn using existing models.
Lastly, try the model on all 12 emotions in the dataset and understand bottlenecks and come up with neural network solutions that can predict with high accuracy.

\section{Link to github code}
$\text{https://github.com/julieeF/CS231N-Project}$

\section{Contributions \& Acknowledgements}

\href{https://www.linkedin.com/in/smandeep/}{\color{blue}Mandeep Singh}:\newline
Mandeep is student at Stanford under SCPD. He has worked at Intel as Design Automation Engineer for 8 years. Prior to joining Intel, he did masters in electrical engineering specializing in analog \& mixed-signal design from SJSU.

Yuan Fang:\newline
Yuan is master's student at Stanford in ICME department. Her interests lies in machine learning \& deep learning.

We would like to thank the CS231N Teaching Staff for guiding us through the project. We also want to thank Google Cloud Platform and Google Colaboratory for providing us free resources to carry out experimentation involved in this work.
{\small
\bibliographystyle{ieee}
\bibliography{egbib}
}

\end{document}